# Attenuation of slow Metal-Insulator-Metal plasmonic waveguides, from Joule absorption to roughness-induced backscattering

Stéphane Coudert [1,*], Guillaume Duchateau [1], Stefan Dilhaire [2], and Philippe Lalanne [3]

[1] Université de Bordeaux-CNRS-CEA, Centre Lasers Intenses et Applications, UMR 5107, 351 Cours de la Libération, 33405 Talence, France
[2] LOMA, Université de Bordeaux-CNRS, UMR 5796, 351 Cours de la Libération, 33405 Talence, France
[3] LP2N, Institut d'Optique Graduate School, CNRS, Univ. Bordeaux, 33400 Talence, France
[*] Correspondence: stephane.coudert@u-bordeaux.fr

**Abstract:** By combining analytical and numerical approaches, we theoretically investigate the effect of fabrication imperfections, e.g. roughness at metal interfaces, on nanometer metal-insulator-metal waveguides supporting slow gap-plasmon modes. Realistic devices with vapor deposition- and chemically-grown metal films are considered. We obtain quantitative predictions for the attenuations induced by absorption and by backscattering, and analytically derive how both attenuations scale with respect to the group velocity. Depending on the material parameters and fabrication quality, roughness-induced backscattering is find to be a significant additional source of attenuation for small group velocities, which is often neglected in the literature.

## 1. Introduction

Much of the impetus for the contemporary interest in hetero-structures combining metals with dielectrics or semiconductors with transverse dimensions much smaller than the wavelength of light is due to the capacity of slow plasmons to confine light and enhance the electromagnetic energy density at the nanoscale [1]. These hetero-structures, such as metal-insulator-metal (MIM) stacks with nanometer dielectric gaps [2], or insulator-metal-insulator (IMI) stacks with thin metal layers [3], are encountered in most of the metamaterial or plasmonic devices. Related applications cover a vast panel, including spectroscopy as surface-enhanced Raman scattering [4], adiabatical nano-focusing [5,6], hot electron generation [7,8], nanoscale localized light sources [9,10,11], near-to-far field effective coupling with nano-antennas [12], and light waveguiding at the subwavelength scale [1,13].

In this paper, we theoretically investigate the effect of fabrication-related disorders on the propagation of light in subwavelength MIM waveguides, in the regime of slow plasmon modes with a small group velocity $v_g$. In this regime, since the field intensity scales with the second power of the group index $n_g = c/v_g$ [14], it is expected that backscattering may become an important factor limiting the propagation of slow plasmons, even somewhat more drastically than for photonic-crystal slow light for which the field intensity "only" scales as $n_g$. Despite this fact, to our knowledge, there are no experimental study on roughness-induced attenuation or loss in slow plasmonic waveguides. Theoretical electromagnetic modeling of small imperfections in narrow geometries is challenging. We are aware of a single earlier theoretical study of the impact of roughness in MIM waveguides for large gap widths (operating away from the slow regime) and very rough surfaces [15].

Another reason for the lack of theoretical results may also arise from the absence of accurate models for the imperfections. An indirect asset of this work is to develop models for analyzing the impact of realistic imperfections on the transport in slow plasmonic waveguides. For that purpose, we consider two main classes of slow waveguides fabrication technologies: 1/ chemically-synthesized metal films or flakes [15,16], for which metal surfaces are essentially atomically flat, except for some occasional monoatomic add-layer defects, and 2/ high-quality polycrystalline metal films deposited by chemical vapor deposition which are composed of 100-nm grains with a small residual roughness, and separated by deep valleys that are remnants of grain boundaries [5,17-24].

In Section 2, we remind the asymptotic behavior of a MIM plasmonic mode, i.e. how the local intensity, the velocity and the shape of the modes change when the gap size tends to zero. We focus on gap plasmon TM modes which do not exhibit cut-off and offer slow propagation as the gap thickness becomes much smaller than the wavelength. This asymptotic analysis will help us understanding how the different sources of attenuation considered in this work compete with each other for small gap sizes.

Sections 3 and 4 are devoted to the modeling and numerical estimations of the impact of imperfection on the transport of slow plasmons. Various sources of attenuation for poly- and mono-crystalline films are considered, local imperfection such as grain boundaries of steep steps, surface imperfections such as roughness at metal insulator interfaces and volumic imperfection such as absorption due to electron-phonon scattering. The absorption losses are evaluated in the case of silver and silicon materials, using optical constants tabulated in [25], for a free-space wavelength of 800 nm. Regarding the roughness induced attenuation in poly-crystalline films, two contributions are considered, grain roughness that we analytically analyze by using a first-Born approximation and grain boundaries that we numerically study by assuming a cuboid well geometry. By summing both contributions, the roughness induced attenuation of the plasmon mode is evaluated with respect to the group velocity. Regarding mono-crystalline metallic film, their influence on the plasmon propagation, in terms of transmission and reflection, has been evaluated through simulations based on the numerical resolution of Maxwell equations.

Our analysis tends to indicate that the plasmon propagation is strongly impacted by both absorption and fabrication imperfections at slow speeds. For state-of-the-art metal roughness of 0.5 nm, the impact of fabrication imperfections can be neglected for group velocities larger than $c/20$, but that it should be considered is the technology gives larger roughness or if one of the two metallic layers is replaced by a semiconductor layer to mitigate material losses [26]. This conclusion with some perspectives are provided in Section 5.

## 2. Slow gap plasmons in MIM waveguides

In this Section, we present an analytical derivation of the properties of slow gap modes guided in MIM hetero-structures with classical electrodynamic models. MIM waveguides support both transverse-electric (TE) and transverse-magnetic (TM) modes [2]. TE-modes may propagate over distances that may even exceed decay lengths observed for TM-polarized waves. Hereafter we are however concerned by gap plasmon TM modes (see Fig. 1). The latter do not exhibit cutoff and offer slow propagations as the gap thickness becomes much smaller than the wavelength.

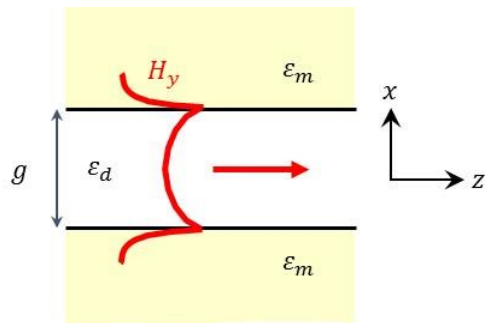

**Figure 1** Symmetric gap plasmon mode (the y-component $H_y$ of the magnetic field component is shown) of MIM waveguides. The mode has no cutoff as the gap thickness $g$ becomes much smaller than the wavelength. The wave vector component $\beta$ is large (and complex if absorption is considered) for small gap thicknesses. The mode is invariant in the $y$-direction. $\varepsilon_m$ and $\varepsilon_d$ denote the dielectric

permittivities of the metal and insulator, respectively. $\text{Re}(\varepsilon_m) < 0$ and $\varepsilon_d > 0$ is assumed to be frequency independent.

Solving the time-harmonic source-free Maxwell's equations for this geometry is a standard procedure, as it looks very similar to the textbook case of dielectric slab waveguides for transverse-magnetic polarization. The magnetic field of the symmetric gap-mode can be written as

$$\begin{cases} H_y(x,z,\beta) = h_0 \left( \exp\left(\gamma_d \left(x - \frac{g}{2}\right)\right) + \exp\left(-\gamma_d \left(x + \frac{g}{2}\right)\right) \right) \exp(i\beta z), & |x| \leq \frac{g}{2}, \\ H_y(x,z,\beta) = h_0 \left(1 + \exp(-\gamma_m g)\right) \exp\left(-\gamma_m \left(|x| - \frac{g}{2}\right)\right) \exp(i\beta z), & |x| > \frac{g}{2}, \end{cases} \quad (1)$$

with $\gamma_\alpha = \pm\sqrt{\beta^2 - \varepsilon_\alpha \frac{\omega^2}{c^2}}$, with $\alpha \equiv m, d$ where $m$ and $d$ stand for metal and dielectric, respectively. The magnetic field amplitude $h_0$ is a constant that will be set afterwards. By using the tangential-field continuity equation for $E_z$ at $x = \pm\frac{g}{2}$, we obtain the transcendental dispersion equation $-\frac{\gamma_m}{\varepsilon_m}(1 + \exp(-\gamma_d g)) = \frac{\gamma_d}{\varepsilon_d}(1 - \exp(-\gamma_d g))$. This equation can be solved for small gaps, i.e. $\frac{\omega}{c} g \ll 1$. Assuming that $\gamma_d g \ll 1$ and $\gamma_d \approx \gamma_m \approx \beta$, which will be verified *a posteriori*, we get

$$\beta g = -\frac{2\varepsilon_d}{\varepsilon_m}, \quad (2)$$

implying that, as the gap-plasmon spatial extension is shrunk by the gap size, the transverse wave-vectors all scale as $\gamma_\alpha \approx 1/g$. Using the expression of the group velocity $v_g = c/n_g = c\left(n + \omega \frac{\partial n}{\partial \omega}\right)^{-1}$ [27] where $n = \beta/k_0$ is the refractive index of the medium and $k_0$ the vacuum wave number of the light, we get from Eq. (2)

$$n_g = \frac{\lambda^2}{g} \frac{\varepsilon_d}{\pi \varepsilon_m^2} \frac{\partial \varepsilon_m}{\partial \lambda}. \quad (3)$$

Just like $\beta$, $n_g$ is a complex number that takes into account the finite propagation length of the gap plasmon due to Ohmic losses. The asymptotic ($\frac{\omega}{c} g \ll 1$) gap-plasmon field distribution takes the following form for $|x| \leq \frac{g}{2}$

$$\begin{cases} H_y(x) = 2h_0 \exp(i\beta z), \\ E_x(x) = 2\beta h_0/(\omega \varepsilon_0 \varepsilon_d) \exp(i\beta z), \\ E_z(x) = i\beta h_0/(\omega \varepsilon_0 \varepsilon_d) \left[\exp(\gamma_d(x - g/2)) - \exp(-\gamma_d(x + g/2))\right] \exp(i\beta z), \end{cases} \quad (4)$$

and for $|x| > \frac{g}{2}$

$$\begin{cases} H_y(x) = 2h_0 \exp(-\gamma_m(|x| - g/2)) \exp(i\beta z), \\ E_x(x) = 2\beta h_0/(\omega \varepsilon_0 \varepsilon_m) \exp(-\gamma_m(|x| - g/2)) \exp(i\beta z), \\ E_z(x) = \frac{x}{|x|} 2i\gamma_m h_0/(\omega \varepsilon_0 \varepsilon_m) \exp(-\gamma_m(|x| - g/2)) \exp(i\beta z). \end{cases} \quad (5)$$

It turns out that the Poynting-vector $z$-component in the gap and in the metals have opposite directions, and that the associated power flow are $-4|h_0|^2 \text{Re}(\omega \varepsilon_0 \varepsilon_m)^{-1}$ and $2|h_0|^2 \text{Re}(\omega \varepsilon_0 \varepsilon_m)^{-1}$, so that the net power flow is $-2|h_0|^2 \text{Re}(\omega \varepsilon_0 \varepsilon_m)^{-1}$. Normalizing this net positive flow to unity at $z = 0$, we get

$$h_0 = (-\omega \varepsilon_0 \, \text{Re} \, \varepsilon_m / 2)^{1/2}, \quad (6)$$

where we have further assumed that $\text{Im} \, \varepsilon_m \ll \text{Re} \, \varepsilon_m$ for the sake of simplicity. We conclude that, for a sufficiently small gap, the amplitude of the magnetic field $h_0$ does not depend on $g$, whereas the electric field amplitudes are inversely proportional to $g$.

In Fig. 2, we consider a gap-plasmon mode with a free-space wavelength $\lambda = 800$ nm propagating in a silver MIM hetero-structure. We compare numerical data obtained with the rigorous coupled-wave analysis [28] for the real and imaginary parts of the group index and the predictions obtained directly from Eq. (3). Despite its simplicity, we find that Eq. (3) is very accurate, especially for small gaps.

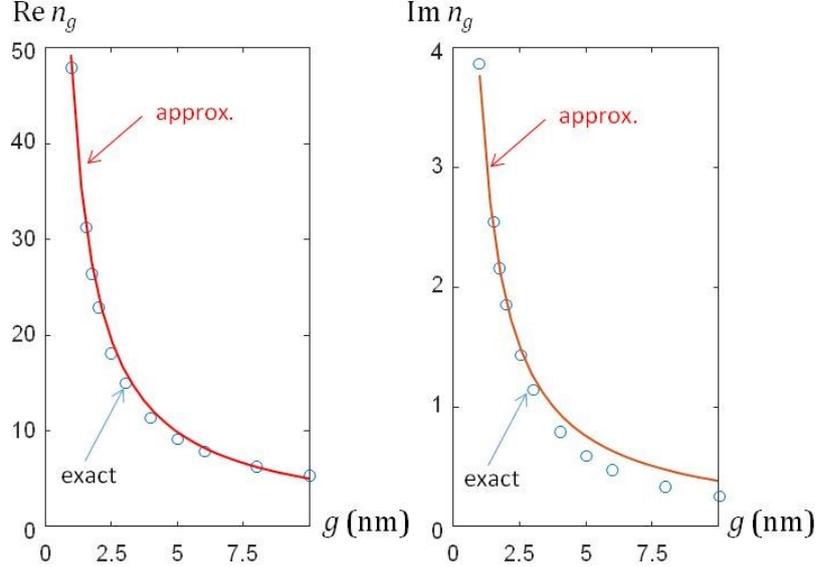

**Figure 2** Test of the accuracy of the simple and approximate Eq. (3) in the limit of the small gaps (i.e. large $\beta$'s) for an MIM waveguide with $\varepsilon_d = 2.25$, $\varepsilon_m = -28.0 + 1.5i$ and $\frac{\partial \varepsilon_m}{\partial \lambda} = -84.23 + 2.63i$ (silver at $\lambda = 800$ nm [25]). Circles are numerical data obtained with the rigorous coupled-wave analysis [28]. The solid red curves correspond to Eq. (3).

## 3. Poly-crystalline films

### 3.1. Model

Physically-grown thin films (e.g. synthetized by vapor deposition) are classically used for the fabrication of MIM cavities [29] or taper waveguides [24]. They are poly-crystalline. Figure 3 sketches the shape of MIMs fabricated with this technology. Each metallic layer is composed of multiple monocrystalline grains with a characteristic size of $\approx 100$ nm, separated by deep valleys. Following [18,19], the surfaces of polycrystalline thin films are characterized by two length scales. The smallest one, close to the atomic scale, is the intrinsic grain roughness. The largest scale is related to the valleys separating the grains. Both imperfections lead to the plasmon attenuation through scattering.

The grain roughness can be described by a Gaussian autocorrelation function with a small standard deviation denoted $\sigma$ and a correlation length denoted $L_c$ [18]. For state-of-the-art films, roughness can be extremely tiny [19], and the standard deviation and the correlation length are in the nanometer range, $\sigma \approx L_c = 0.5 - 1$ nm. In our analysis, we separately consider the effects of valleys and grain roughness. The former is analyzed with fully numerical methods by solving Maxwell's equations. The latter is analyzed by using a first-order Born approximation with local-field corrections.

Throughout the manuscript, the optical properties of such films are evaluated below, including backscattering and absorption, in order to evaluate their influence on the expected plasmonic properties of MIM hetero-structures. Throughout the manuscript, the free-space wavelength is 800 nm, the metal is silver with a relative permittivity $\varepsilon_m = -28 + i1.5$ taken from tabulated data in [25], and the gap material is assumed to be SiO$_2$ ($\varepsilon_d = 2.25$).

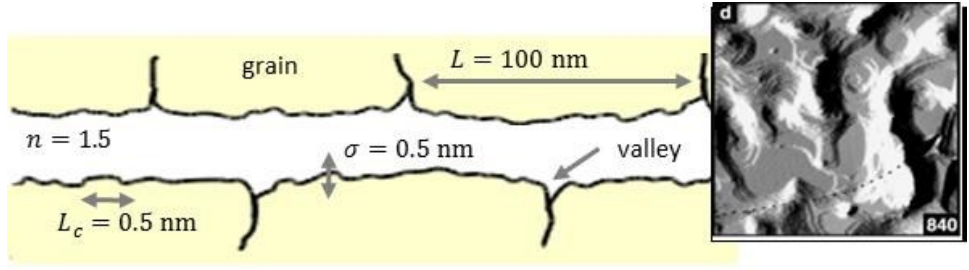

**Figure 3** Schematic representations of roughness models for metal(Ag)/insulator $\varepsilon_d = 2.25$/metal(Ag) polycrystalline film obtained by chemical vapor deposition. The film is composed of multiple monocrystalline grains with a characteristic size of $\approx 100$-nm, an intrinsic roughness $\sigma$ and an autocorrelation length $L_c$. The grains are separated by deep valleys, which reflect and absorb the propagating plasmons.. The right inset shows the profile of a thin gold film measured with an AFM [19].

### 3.1.1 Reflection induced by grain boundaries

To evaluate the impact of grain boundaries, we solve Maxwell's equation with a fully-vectorial computational method, and for the sake of simplicity we model the valley between two adjacent grains [17,18] by a cuboid groove introduced into an MIM waveguide with flat interfaces, see Fig. 4(a).

In our simulation, consistently with observations performed with a scanning electron microscope, we assume that the groove width and depth are equal and denoted by $H$ as shown in fig. 4. Two characteristic values, $H = 2$ nm and $H = 4$ nm are considered. The simulations are performed using an in-house aperiodic Fourier Modal method (a-FMM) [30]. The latter solves Maxwell's equations using Fourier expansion techniques in the MIM transverse directions and a scattering-matrix formalism in the mode-propagation direction. It has been benchmarked for plasmonic geometries composed of MIM and grooves, similar to the present one, and the comparison with other established methods of electromagnetism, operating either in the temporal or frequency domains, has evidenced its high performance [31].

For ultra-narrow gaps and valleys, nonlocal effects may take place in the metal due to strong field gradients [32]. To evaluate their potential impact on our classical numerical predictions, we replace the actual metal by a composite material, comprising a thin dielectric layer at the metal surfaces and a local metal modeled by the permittivity $\varepsilon_m$. This thin layer is represented by thick red lines in the left panel of Fig. 4(a). This approach makes possible the quantitative analysis of nonlocal effects in complex plasmonic devices with numerical treatments of classical electrodynamics [32], and bypasses complicated first-principles methods [33]. For the thickness $\Delta d$ and relative permittivity $\varepsilon_t$ of the thin layer, we choose $\Delta d = 0.1$ nm and $\varepsilon_t = q_L \Delta d\, \varepsilon_b \varepsilon_m / (\varepsilon_m - \varepsilon_b)$, with $q_L = \sqrt{\omega_p^2/\varepsilon_\infty - \omega(\omega + i\gamma)}/v_{nl}$. Here, $\omega_p$ and $\gamma$ denote the metal damping and plasma frequency, and the $v_{nl}$ factor, proportional to the Fermi velocity, measures the degree of nonlocality. For the simulations, we take $v_{nl} = 1.27\,10^6$ m/s [32], $\varepsilon_\infty = 5$, $2\pi c/\omega_p = 138$ nm, $\gamma = 0.0063\omega_p$, so that the Drude permittivity $\varepsilon_m = \varepsilon_\infty - \omega_p^2/\omega(\omega + i\gamma)$ correctly fits the silver-permittivity data in [25] from 400 nm to 1 µm. As shown in [32], with those values, the emergence of nonlocality effects can be reproduced to a high degree of precision in a variety of plasmonic systems. Note that the thin layers are always placed at the structural interfaces in the metal regions, in such a way that the gap width remains the same as in the original geometry. Also note that the quantum corrections we observe are not significant even for the smallest gap dimensions we consider hereafter, and that electron tunneling is completely negligible too.

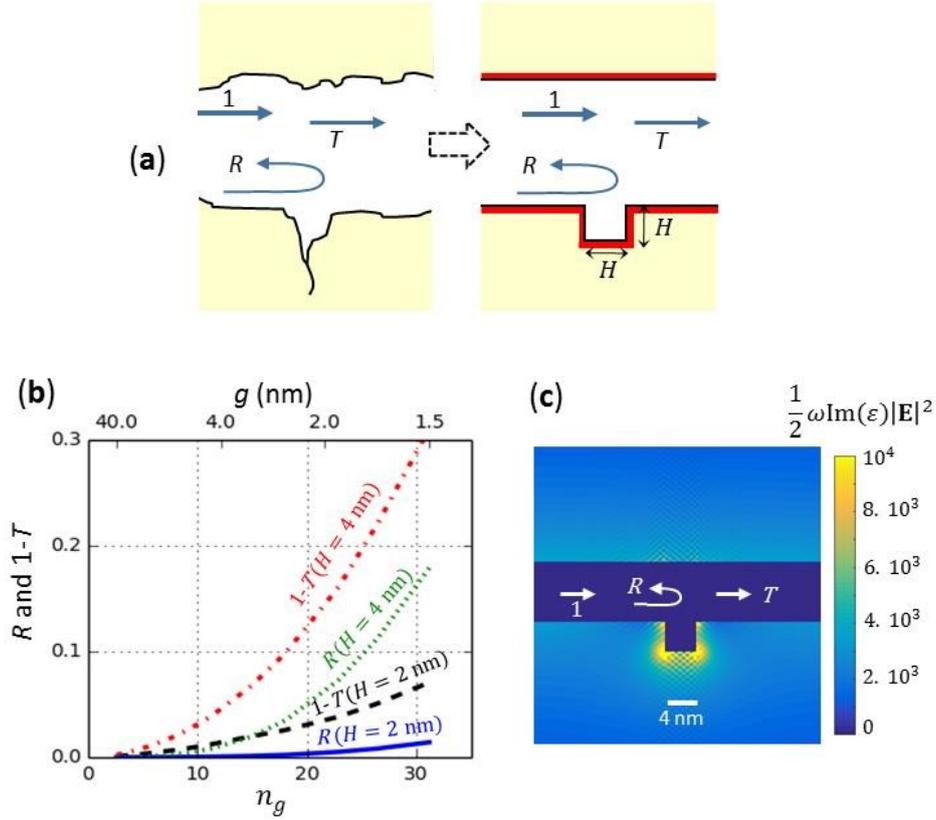

**Figure 4** Modelisation of the reflectance and absorbance of the valleys at the grain boundaries of polycrystalline films. (a) The realistic valleys (left) are represented as square groves (right), with identical depth and width $H$. The thick red lines in the left panel depict the thin dielectric layer added in the simulation to model nonlocal effects at the metal interfaces, see text. (b) Reflectance $R$ and extinction $1 - T - R$ due to the scattering of a gap plasmon mode by a single square groove as a function of the plasmon group index $n_g$ (lower horizontal axis) or the gap width $g$ (upper axis). Note that $1 - T - R \neq 0$, highlighting a strong metal absorption at the groove. (c) Absorbed power density $\frac{1}{2}\omega Im(\varepsilon)|\mathbf{E}|^2$, for an incident gap plasmon with a unitary power flow and for $g = 4$ nm and $H = 2$ nm. Note that the absorbed power density of the incident gap plasmon propagating in the flat MIM sections is not revealed with the linear colorbar used.

Figure 4(b) presents the numerical predictions obtained for the reflectance $R$ and the transmittance $T$ of a gap plasmon for two groove dimensions, as a function of the plasmon group index (lower horizontal axis) or gap size (upper one). Note that we plot the extinction $1 - T$, rather than the transmittance $T$, in order to highlight that the absorption $1 - T - R$ is not equal to zero. When the gap width decreases, the local amplitude of the plasmon intensity increases due to both the slowing down and the confinement, and as expected, the impact of the groove on the fraction of reflected power increases monotonically.

We observe that the dependence of the backscattered power with respect to the groove dimension is strongly nonlinear; the power is approximately increased by a factor 10 when the groove size is only increased by a factor 2. Also, comparatively, the increase of the backscattered power with the groove dimension is much faster than the increase of the extinction for a same group index (×10 versus ×4 for $n_g = 30$). The absorption (or the Joule loss) can be estimated by evaluating $1 - T - R$; It is far from being negligible and actually is almost one half of the reflectance for any gap width. Figure 4(c) shows the absorbed power density $\frac{1}{2}\omega Im(\varepsilon)|\mathbf{E}|^2$, computed for an incident gap plasmon with a unitary power flow and for a gap width $g = 4$ nm and a groove dimension $H = 2$ nm. It shows that absorption dominantly takes places in the bottom of the groove and on the vertical walls.

### 3.1.2 Backscattering due to grain roughness

Due to its high permittivity contrast and deep subwavelength dimensions, model grain roughness is challenging to analyze with fully-vectorial approaches. Thus we adopt a fully analytical formalism developed in [34]. The formalism relies on perturbation theory and local field corrections, which allow for quantitative predictions [35].

An imperfection at an interface, a metal dip ($\Delta l < 0$) or a metal hump ($\Delta l > 0$) relative to the perfectly flat interface described by the permeability distribution $\varepsilon(x, z)$ ($x$ denotes the dimension orthogonal to the flat interface and $z$ the direction of propagation of the plasmon mode tangential to the flat interface), scatters the incident gap plasmon. The Maxwell's equations for the total electric field $\mathbf{E}_T$ and magnetic field $\mathbf{H}_T$ at the angular frequency $\omega$ read as

$$\begin{cases} \nabla \times \mathbf{E}_T = i\omega\mu_0 \mathbf{H}_T, \\ \nabla \times \mathbf{H}_T = -i\omega\varepsilon(x)\mathbf{E}_T - i\omega\Delta\varepsilon(x,z)\mathbf{E}_T \end{cases} \quad (7)$$

in the two-dimensional Cartesian coordinate $(x, z)$ system. Equation (7) shows that the imperfections act as current sources proportional to the perturbation $\Delta\varepsilon(x, z)$ of the dielectric permittivity and the total electric field $\mathbf{E}(x, z)$ at the position of the imperfection.

Hereafter, we will consider the backscattering induced by a single interface, the one located at $x = 0$, see Fig. 5. The effect of the other interface can be treated identically and will result in an attenuation of reflectance twice larger. Using mode orthogonality, it can be shown that the modal reflection coefficient of the incident gap plasmon is given by

$$r = -\frac{\omega}{4} \varepsilon_0 \Delta\varepsilon \int \Delta l(z) \, \mathbf{E}(x = 0, z) \cdot \mathbf{E}_T(x = 0, z) dz, \quad (8)$$

where $\mathbf{E}(x, z)$ is the mode distribution of a gap plasmon defined by Eqs. (4) and (5) normalized such that its Pointing flux is unity, implying that $h_0$ satisfies $4\omega h_0^2/(\varepsilon_0 \beta^2 c^2) = 1$. To derive Eq. (8) we have assumed that the fields do not depend on $x$ at the scale of the roughness $\Delta l$.

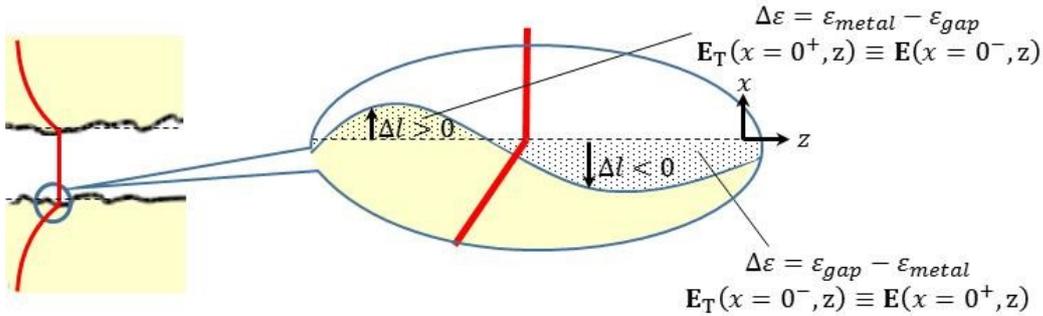

**Figure 5** Modified first-order Born approximation leading to Eq. 9. The shape of the unperturbed plasmon mode is shown in red and the dashed horizontal lines represent the interfaces of the unperturbed MIM. The dotted areas represent the perturbed regions for which we highlight the local field corrections used to estimate the actual perturbed electric field $\mathbf{E}_T(x, z)$ from the unperturbed plasmon-mode electric field $\mathbf{E}(x, z)$.

Equation (8) is exact, but it contains the unknown electric field distribution $\mathbf{E}_T(x, z)$. An approximate expression for $\mathbf{E}_T(x, z)$ can be obtained with perturbation theory, assuming that the plasmon damping is due to roughness and significantly larger than the plasmon decay length. Classical first-Born approximation amounts to assuming that the perturbed field $\mathbf{E}_T(x, z)$ can be simply replaced by the unperturbed field $\mathbf{E}(x, z)$ in Eq. (8). As shown in the appendix in [34], this simple approach can be much improved with local field corrections. Referring to Fig. 2 and $\Delta l > 0$, the perturbed field in the dotted region for $x = 0^+$ is in the metal, so that a better approximation for $\mathbf{E}_T(0^+, z)$ for $\Delta l > 0$ is not $\mathbf{E}_T(0^+, z) \equiv \mathbf{E}(0^+, z)$ but rather $\mathbf{E}_T(0^+, z) \equiv \mathbf{E}(0^-, z)$. Similarly for $\Delta l < 0$, $\mathbf{E}_T(0^-, z) \equiv \mathbf{E}(0^+, z)$.

The mean reflectance $\langle R \rangle = \langle rr^* \rangle$ of the perturbed section of length $L$ is then easily derived and using $\langle \Delta l \rangle = 0$, we get

$$\langle R \rangle = 2\left(\frac{\omega}{4}\right)^2 \Delta\varepsilon_r^2 \sigma^2 \int_0^L \int_0^L \varepsilon_0^2 \; \Gamma(z_1, z_2) \, \theta(z_1) \, \theta^*(z_2) \, dz_1 dz_2, \tag{9}$$

where $\Gamma(z_1, z_2) = \frac{1}{\sigma^2} < \Delta l(z_1) \; \Delta l(z_2) >$ is the autocorrelation function and $\theta(z)$ is given by

$$\theta(z) = E_z^2(z) + D_x^2/(\varepsilon_d \varepsilon_m). \tag{10}$$

The factor two in Eq. (9) accounts for the response of the two interfaces of the MIM, whose roughness are assumed to be independent. In Eq. (10), which directly results from the local field corrections, $E_z$ is the $z$-component of the gap-mode electric field and $D_x$ is the $x$-component of the gap-mode displacement vector.

Using a realistic white Gaussian spectrum for the roughness with an autocorrelation length $L_c$ and a root mean square height $\sigma$ (rms) as in [15], we obtain $\langle R \rangle = 2\left(\frac{\omega}{4}\right)^2 \Delta\varepsilon_r^2 \varepsilon_0^2 \, \sigma^2 \, |\theta|^2 \, L \, L_c \, \sqrt{\pi} \, e^{-\beta^2 L_c^2}$. We emphasize that this development is valid if imperfections are significantly smaller than the plasmon decay length, leading us to consider that $\beta L_c \ll 1$:

$$\langle R \rangle = 2\left(\frac{\omega}{4}\right)^2 \Delta\varepsilon_r^2 \varepsilon_0^2 \, \sigma^2 \, |\theta|^2 \, L \, L_c \, \sqrt{\pi} \,, \tag{11}$$

which is valid if $\langle R \rangle \ll 1$. The attenuation coefficient due to the roughness, $\alpha_{\text{rough}} = \langle R \rangle / L$ at small reflectance, then reads as

$$\alpha_{\text{rough}} = 2\left(\frac{\omega}{4}\right)^2 \Delta\varepsilon_r^2 \varepsilon_0^2 \, \sigma^2 \, |\theta|^2 \, L_c \, \sqrt{\pi} \,, \tag{12}$$

It is interesting to consider the behavior of $\alpha_{\text{rough}}$ with respect to the group index in the limit of very small gaps, when the wave number and the group index are proportional to the inverse of the gap width, see Eqs. (2) and (3). Since the amplitude of the electric field is proportional to the inverse of the gap width, $\theta$ is proportional to $g^{-2}$, meaning that $\alpha_{\text{rough}}$ is proportional to $g^{-4}$ (or $n_g^4$).

**3.2 Discussion**

Figure 6 summarizes the results of our calculations for the attenuation coefficient $\alpha$ of slow gap plasmons propagating in a polycrystalline MIM hetero-structure. The results are plotted as a function of either the plasmon group index or the gap size. We use the standard definition for the attenuation with an exponential decay of the intensity $\exp -\alpha z$. Several curves are represented.

All results are obtained for a silver MIM for $\lambda = 800$ nm. The blue solid curve shows the attenuation coefficient due to Joule absorption.

The attenuation rate $\alpha_{\text{rough}}$ due to roughness-induced backscattering is shown with the dashed curves for $L_c = \sigma = 0.5$ nm and $L_c = \sigma = 1.0$ nm. First, we note that $\alpha_{\text{rough}}$ strongly depends on the roughness size. This is consistent with Eq. (12) that evidences that $\alpha_{\text{rough}}$ varies as $\sigma^2 L_c$ for small values of $\beta L_c$. If we further assume that $\sigma$ and $L_c$ are proportionally at small roughness, $\alpha_{\text{rough}}$ is expected to scale with the third power of the characteristic roughness parameters. The second important observation is that $\alpha_{\text{rough}}$ rapidly increase with $n_g$, and is proportional to $n_g^4$ for group indices larger than 10. This is again consistent with Eq. (12); $\theta$ being proportional to the squared value of the electric field at the interface, i.e. proportional to $n_g^2$, $|\theta|^2$ effectively scales as $n_g^4$.

The attenuation rate $\alpha_{\text{val}}$ induced by plasmon scattering at the grain boundaries is shown for two groove dimensions, $H = 2$ nm and $H = 4$, with the thin-solid curves. These predictions assume that the separation distance between two independent valleys is equal to the grain size $L = 100$ nm, so that $\alpha_{\text{val}} = \frac{1-T}{L}$, with $T$ given in Fig. 4. The valley attenuations exhibit identical $n_g^2$ scalings, which are more sensitive than the $n_g$-scaling of the Joule attenuation, but remains less drastic than the $n_g^4$-scaling of the roughness-induced backscattering.

For the state-of-the-art roughness parameters [17,18] used for Fig. 6, the damping due to Joule absorption remains significantly larger than the attenuation due to backscattering at the grain-boundaries. It is also larger than the roughness-induced attenuation, except for tiny gap widths of $\approx 1$-3 nm, for which absorption and roughness have comparable impacts. This evidences that, for poly-crystalline films, much attention must be paid to the choice of a suitable technology to avoid that roughness perturbs the transport of slow plasmons, or the lifetime of plasmon resonances in MIM cavities which gap size is smaller than 2 nm.

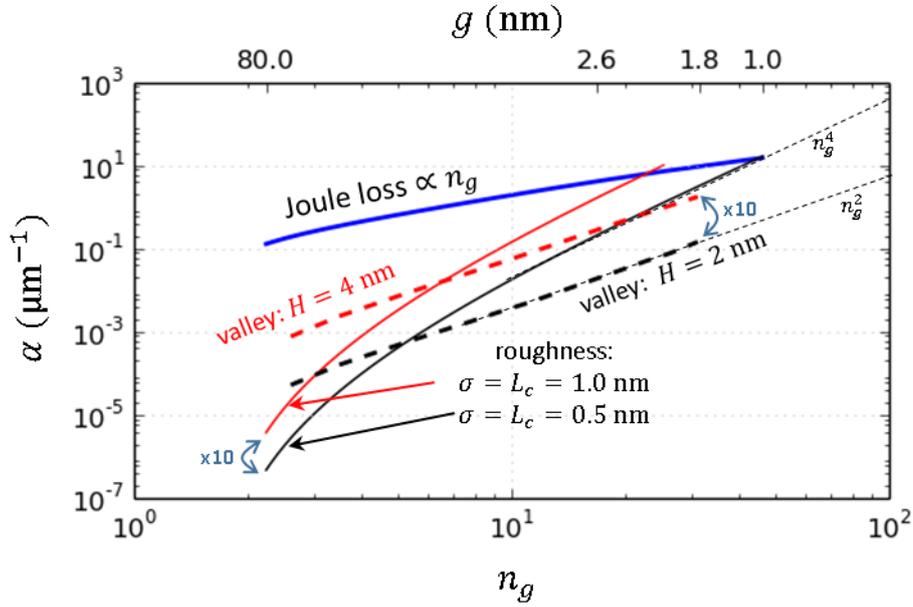

**Figure 6** Extinction rates of Ag/insulator/Ag slow waveguides for a grain size $L = 100$ nm. Solid-black and –red thin curves: attenuation rate due to grain roughness for two values of $L_c$ and $\sigma$. Dashed-black and -red curves: attenuation rate due to the valleys for two values of $H$. The rates are computed as the product of the extinction $(1 - T)$ due to a single valley by the valley density $1/L$ along the direction of propagation. The curves are directly obtained from the numerical results shown in Fig. 4. Blue solid thick curve: attenuation due to metal absorption. The waveguide is composed of a dielectric gap ($\varepsilon_d = 2.25$) inserted between two silver films ($\varepsilon_m = -28 + i1.5$ at $\lambda = 800$ nm). Up and down thin dashed black lines represent respectively the scaling behavior $n_g^4$ and $n_g^2$ of the attenuation rates.

## 4. Mono-crystalline films

For chemically grown thin films, such as flakes or nanoparticles (core-shell, nanocubes on substrate), the interfaces are basically atomically flat. However, atomic steps might take place randomly. For MIM waveguides offering $n_g > 10$, the height of mono-atomic-step defects becomes comparable to the gap width. This raises the question of to what extent a mono-atomic step may impact the attenuation of slow plasmons propagating in MIM waveguides. To answer the question, we perform numerical computations using the same approach as in Section 3, modeling the mono-atomic ad-layer defect as a permittivity step of height $h = 0.5$ nm and further considering the potential effect of nonlocal effect by incorporating a thin isotropic layer, which is schematically represented in Fig. 7(a).

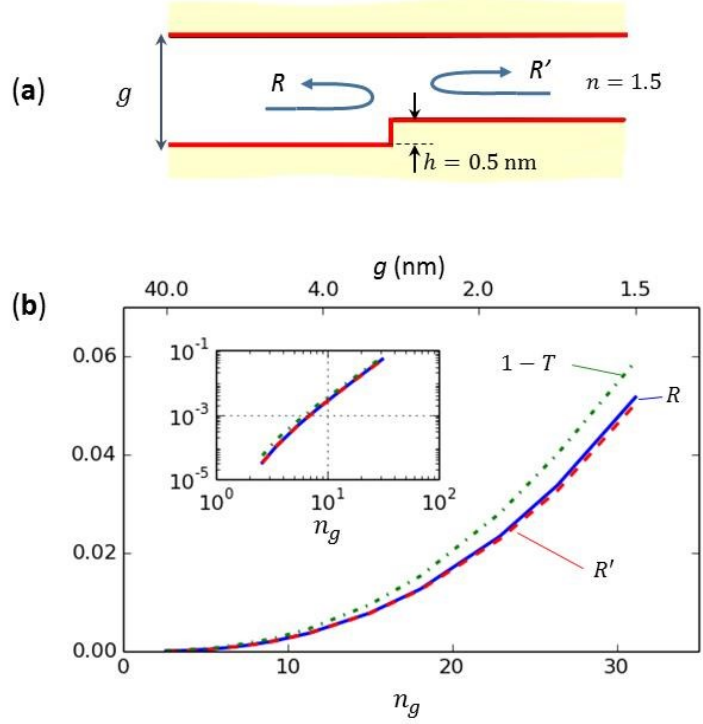

**Figure 7** Impact of mono-atomic-step defects on chemically synthesized Ag/insulator/Ag slow waveguides. **(a)** Mono-atomic-step defect modeled by a permittivity wall. The thick red lines depict the thin dielectric layer added in the simulation to model nonlocal effects. **(b)** $R$ denotes the reflectance for plasmons incident from the large gap, while $R'$ denotes the reflection coefficient for a plasmon coming from the thin gap. The full-wave computational results are obtained for a 0.5-nm height corresponding to a metal monolayer. Inset: log-log representation, highlighting the $n_g^3$ scaling of $R$, $R'$ and $1-T$. The waveguide is composed of a dielectric gap ($\varepsilon_d = 2.25$) inserted between two Ag films ($\varepsilon_m = -28 + i1.5$ at $\lambda = 800$ nm).

The step defines an interface between two MIM waveguides with slightly different gap thicknesses, and accordingly two intensity reflection coefficients, denoted $R$ and $R'$. Figure 7(b) presents the computational results for $R$ and $R'$ as a function of the group index $n_g$ of the plasmon propagating in the waveguide with the largest gap $g$. We also plot the extinction $1-T$ with the green dashed-dotted curve. It is noticeable that $R + T \approx 1$, implying that the Joule absorption at the step is small (< 0.01) for all speeds. Owing to reciprocity, we know that the transmission does not depend on the direction of the incident plasmon propagation, so that we need to plot only one transmission curve. The three coefficients, $R$, $R'$ and $1-T$, are plotted with a log-log scale in the inset, highlighting their $n_g^3$ scaling. This scaling law can be understood from geometrics considerations, assuming that the reflection linearly increases with the aspect ratio $\frac{h}{g}$, i.e. with $n_g$, see Eq. (3), and noting that the intensity of normalized plasmon mode increases linearly with $n_g^2$ in the insulator.

In contrast to the deep valleys that occur every 100 nm, mono-atomic-step defect in chemically-synthesized films are rare, and it does not make sense to extrapolate an attenuation rate from the numerical prediction of the extinction $1-T$.

## 5. Conclusions

Benefiting from recent works on the characterization of the roughness of thin metal films, we have proposed a physical model for theoretically analyzing the impact of fabrication imperfections on the attenuation of slow gap plasmons in MIM waveguides. State-of-the-art mono- and poly-crystalline technologies have been analyzed.

The study that combines analytical developments and computational results obtained with fully-vectorial electromagnetic tools provides a good understanding of the various effect that impact the attenuation. It tends to indicate that the plasmon propagation is severely impacted by both absorption and fabrication imperfections at slow speeds.

Undoubtedly, the preferred technology for slow MIM waveguides is chemically-synthesized films.

For poly-crystalline films fabricated with physically growth and deposition, our results evidence that backscattering induced by grain roughness dominates over backscattering induced by grain boundary and that backscattering is much smaller than absorption at small group indices. However, the conclusions are more balanced for slow plasmons in tiny gaps. Since absorption scales with $n_g$ whereas roughness-induced backscattering scales with $n_g^4$, the impact of fabrication imperfections at small group velocities should be considered. Actually, for tiny gaps supporting slow modes with $n_g \approx 20$ ($g \approx 2 - 3$ nm), absorption and roughness attenuations become comparable. We recall that this conclusion hold for silver MIM operating in the near-infrared ($\lambda = 800$ nm) and for state-of-the-art technologies with a surface roughness of $\sigma = 1$ nm (rms).

Moreover, since backscattering rapidly increases with the roughness of the surface ($\propto \sigma^2$ as predicted by our analytical model), it is expected that for film depositions providing surface roughness of a few nanometers rms, backscattering may become the dominant attenuation factor even for moderately large refractive indices of 10. Much care should be taken to optimize the growth or the deposition of thin metallic films to effectively benefit from slow plasmon technologies in MIM waveguides.

## Acknowledgements

P.L. is pleased to acknowledge the support from CNRS and Jean-Paul Hugonin for a long standing collaboration on the development of Fourier Modal numerical methods. S.C. and G.D. thank the University of Bordeaux for providing a PhD grant.